\documentclass[twocolumn]{aastex63}
\usepackage{amsmath,amssymb,graphicx,longtable,booktabs,color,CJK,natbib,threeparttable,soul,bm}

\shorttitle{CG of GC Systems in M87 and M49}
\shortauthors{Wu et al.}

\begin{document}
\begin{CJK*}{UTF8}{gbsn}

\title{The Color Gradients of the Globular Cluster Systems in M87 and M49.}

\correspondingauthor{Chengze Liu}
\email{czliu@sjtu.edu.cn}

\author{Yiming Wu}
\affiliation{Department of Astronomy, School of Physics and Astronomy, and Shanghai Key Laboratory for Particle Physics and Cosmology, Shanghai Jiao Tong University, Shanghai 200240, China}

\author[0000-0002-4718-3428]{Chengze Liu}
\affiliation{Department of Astronomy, School of Physics and Astronomy, and Shanghai Key Laboratory for Particle Physics and Cosmology, Shanghai Jiao Tong University, Shanghai 200240, China}

\author[0000-0002-2073-2781]{Eric W. Peng}
\affiliation{Department of Astronomy, Peking University, Beijing 100871, China}
\affiliation{Kavli Institute for Astronomy and Astrophysics, Peking University, Beijing 100871, China}

\author[0000-0001-6333-599X]{Youkyung Ko}
\affiliation{Korea Astronomy and Space Science Institute, 776 Daedeok-daero, Yuseong-Gu, Daejeon 34055, Korea}

\author[0000-0003-1184-8114]{Patrick C\^ot\'e}
\affiliation{Herzberg Astronomy and Astrophysics Research Centre, National Research Council of Canada, 5071 W. Saanich Road, Victoria, BC, V9E 2E7, Canada}

\author{Rashi Jain}
\affiliation{Universit\'e de Strasbourg, CNRS, Observatoire astronomique de Strasbourg, UMR7550, F-67000, Strasbourg, France}

\author[0000-0002-8224-1128]{Laura Ferrarese}
\affiliation{Herzberg Astronomy and Astrophysics Research Centre, National Research Council of Canada, 5071 W. Saanich Road, Victoria, BC, V9E 2E7, Canada}

\author[0000-0003-3997-4606]{Xiaohu Yang}
\affiliation{Department of Astronomy, School of Physics and Astronomy, and Shanghai Key Laboratory for Particle Physics and Cosmology, Shanghai Jiao Tong University, Shanghai 200240, China}

\author[0000-0002-7214-8296]{Ariane Lan\c{c}on}
\affiliation{Universit\'e de Strasbourg, CNRS, Observatoire astronomique de Strasbourg, UMR7550, F-67000, Strasbourg, France}

\author[0000-0003-0350-7061]{Thomas Puzia}
\affiliation{Instituto de Astrofsica, Pontificia Universidad Cat\'olica de Chile, Av. Vicu\~na Mackenna 4860, 7820436 Macul, Santiago, Chile}

\author[0000-0002-5049-4390]{Sungsoon Lim}
\affiliation{University of Tampa, 401 West Kennedy Boulevard, Tampa, FL 33606, USA}

\begin{abstract}
Combining data from the ACS Virgo Cluster Survey (ACSVCS) and the Next Generation Virgo cluster Survey (NGVS), we extend previous studies of color gradients of the globular cluster (GC) systems of the two most massive galaxies in the Virgo cluster, M87 and M49, to radii of $\sim 15~R_e$ ($\sim 200$ kpc for M87 and $\sim 250$ kpc for M49). We find significant negative color gradients, i.e.,  becoming bluer with increasing distance, out to these large radii. The gradients are driven mainly by the outwards decrease of the ratio of red to blue GC numbers. The color gradients are also detected out to $\sim 15~R_e$ in the red and blue sub-populations of GCs taken separately. In addition, we find a negative color gradient when we consider the satellite low-mass elliptical galaxies as a system, i.e., the satellite galaxies closer to the center of the host galaxy usually have redder color indices, both for their stars and GCs. According to the ``two phase" formation scenario of massive early-type galaxies, the host galaxy accretes stars and GCs from low-mass satellite galaxies in the second phase. So the accreted GC system naturally inherits the negative color gradient present in the satellite population. This can explain why the color gradient of the GC system can still be observed at large radii after multiple minor mergers.
\end{abstract}



\section{Introduction}
\label{sec:intro}

Most, but not all, galaxies show negative color gradients, i.e., the stellar components are redder in galaxy core region than those in the outskirts \citep[e.g.][]{2009RAA_9_1119Liu, 2011MNRAS_416_1983Roediger, 2012MNRAS_425_841Loubser, 2016A+A_593_84Kennedy, 2017MNRAS_465_4572Zheng, 2018RAA_18_143Liang, 2019ApJ_877_103Suess, 2020MNRAS_495_2894DominguezSanchez}. The color gradients of most of the early-type galaxies (ETGs) are mainly due to the radial variation of metallicity, while the origin of color gradients for late-type galaxies are much more complicated \citep[e.g.,][]{2005AJ_129_2628Liu, 2011MNRAS_416_1996Roediger, 2012ApJ_758_41Roediger, 2013ApJL_764_20Mihos, 2017MNRAS_466_4731Goddard, 2020MNTAS_495_3387Peterken}. The steepness and sign of color gradients are closely related to the formation and assembly history of galaxies, e.g., the monolithic collapse formation scenario predicts the existence of negative color gradients \citep[e.g.][]{2009ApJL_699_178Naab, 2010MNRAS_407_1347Pipino, 2016ApJ_828_27Nelson, 2017MNRAS_471_3856Taylor} while the galaxy assembly through mergers (either major or minor) tends to flatten existing gradients \citep[e.g.,][]{2004MNRAS_347_740Kobayashi, 2009A+A_499_427DiMatteo, 2012MNRAS_421_2478Tortora, 2019ApJ_880_111Oyarzun}.

According to the standard cosmological model, massive galaxies are believed to form hierarchically. In this framework, the formation of massive ETGs have been well-described by a ``two phase" formation scenario \citep[e.g.,][]{1998ApJ_501_554Cote, 2010ApJ_725_2312Oser, 2012ApJ_754_115Johansson,2013ApJL_768_28Huang, 2017ApSS_362_206Modak, 2021MNRAS_504_4923Dolfi}.  In this scenario, the stars formed \textit{in situ} though the dissipative collapse of gas at an early stage, followed by a second phase of successive minor mergers, which accrete stars (formed \textit{ex situ}) from smaller satellite galaxies. The fraction of \textit{ex situ} stars increases with radius, and dominates in the outskirts of the galaxy \citep{2021MNRAS_507_3089Davison, 2021A+A_647_95Pulsoni}. These accreted stars increase the masses and sizes of the main galaxies \citep[e.g.,][]{2009ApJL_699_178Naab, 2012ApJ_744_63Oser, 2019MNRAS_487_318Karademir} and modify the color (stellar population) profiles, especially in the galaxy outskirts \citep[e.g.][]{2009A+A_499_427DiMatteo, 2011ApJ_735_18Guo, 2012MNRAS_421_2478Tortora, 2020MNRAS_491_3562Zibetti}. \citet{2019MNRAS_487_318Karademir} investigated how the orbit configurations influence the mass distribution of the merger remnant and the disrupted satellite galaxy. They found that the streams are usually the result of nearly circular infall while the shell structures are made from radial satellite infall. Different merger events modify the mass and stellar population distribution in different ways. The Feedback in Realistic Environments (FIRE) and FIRE-2 simulation \citep{2014MNRAS_445_581Hopkins, 2018MNRAS_480_800Hopkins} investigated how the feedbacks influence the mass growth, the  morphology, the quenching and other properties of galaxies. In particular, they showed that the feedback from rapid infall and strong outflow can stir the distribution of gas and stars, and flatten the metallicity gradient \citep{2016MNRAS_456_2140Ma, 2017MNRAS_466_4780Ma}.

The investigation of color/metallicity profiles and the gradients, especially at large radii, can give hints on the formation and assembly history of massive ETGs. However, the low surface brightness of galaxies' outskirts makes it difficult to study their profiles at very large radii, even when stacking high-quality deep images \citep[][and references therein]{2018MNRAS_475_3348Huang}.

As one of the oldest stellar systems in the Universe, globular clusters (GCs) are tightly tied to the formation and assembly history of their host galaxies. In addition, GCs are much easier to observe since they are compact and much brighter than individual stars \citep[e.g.,][]{2009ApJ_703_939Harris, 2011MNRAS_413_2943Forbes, 2015MNRAS_451_2625Pastorello, 2019MNRAS_482_950Forte}. So the GC system is an alternative and unique probe of formation and assembly history of their host galaxy. 

Using GC systems as probes, \citet{1998ApJ_501_554Cote} investigated the formation and evolution of massive ETGs. They originally proposed that the red (metal-rich) GCs are the intrinsic population and the blue (metal-poor) GCs are accreted from other galaxies, and the number ratio of blue and red GCs can estimate the merger history of the massive ETG. The simulation studies predicted that the blue GCs formed at early time ($z \sim 4$) in low-mass halos while the red GCs formed later ($z \sim 2$) in more massive halos \citep{2018MNRAS_480_2343Choksi, 2019MNRAS_482_4528ElBadry, 2019MNRAS_486_5838ReinaCampos}. With the hierarchical growth, massive galaxies will naturally have red and blue GC populations.

Based on ACS Virgo Cluster Survey \citep[ACSVCS,][]{2004ApJS_153_223Cote} and ACS Fornax Cluster Survey \citep[ACSFCS,][]{2007ApJS_169_213Jordan} data, \citet{2011ApJ_728_116Liu} studied the color and metallicity gradients of GC systems of Virgo and Fornax ETGs. They found that the GC systems in most ETGs have negative color/metallicity gradients, which is consistent with the results of other studies \citep[e.g.,][]{2009ApJ_699_254Harris, 2009ApJ_703_939Harris, 2011MNRAS_416_155Faifer, 2015MNRAS_451_2625Pastorello, 2019ApJ_872_202Ko}. However, the field of view of HST/ACS is relatively small with $202 \arcsec \times 202 \arcsec$, which can only cover $\sim R_e$ for the most massive ETGs in the Virgo cluster. 

The Next Generation Virgo Cluster Survey \citep[NGVS,][]{2012ApJS_200_4Ferrarese} is a deep, multi-wavelength imaging survey that covers the whole Virgo cluster ($\sim 104$ deg$^2$). Based on the excellent NGVS images, several papers have been published to investigate the GCs in the Virgo cluster \citep{2014ApJ_794_103Durrell, 2014ApJS_210_4Munoz, 2014ApJ_792_59Zhu, 2015ApJ_802_30Zhang, 2016ApJS_227_12Powalka, 2016ApJL_829_5Powalka,  2016ApJ_822_51Toloba, 2017ApJ_844_104Powalka, 2018ApJ_864_36Longobardi, 2018ApJ_858_37Zhang, 2018ApJ_856_84Powalka, 2019ApJ_885_145Sun, 2021ApJ_915_83Taylor}. With the combination of ACSVCS (high spatial resolution) and NGVS (wide field of view) data, we have a unique chance to study the color gradients of GC systems of massive ETGs at very large range in radii (from the center to the edge of the galaxy). 

In this study, we will focus on the two most massive ETGs in the Virgo cluster, M87 (VCC1316) and M49 (VCC1226), which have likely undergone many major/minor mergers in their assembly history. We will introduce the data and GC sample selection in Section~\ref{sec:data}. In Section~\ref{sec:results}, we present our results and the discussion, and in Section~\ref{sec:summary} we give a brief summary of our findings. We adopt an effective radius $R_e = 171.71' \simeq 13.74$ kpc for M87 and $R_e = 209.15' \simeq 16.73$ kpc for M49 \citep{2006ApJS_164_334Ferrarese} in this study. 

\section{Data and GC Sample}
\label{sec:data}

\begin{figure}
\epsscale{1.1}
\plotone{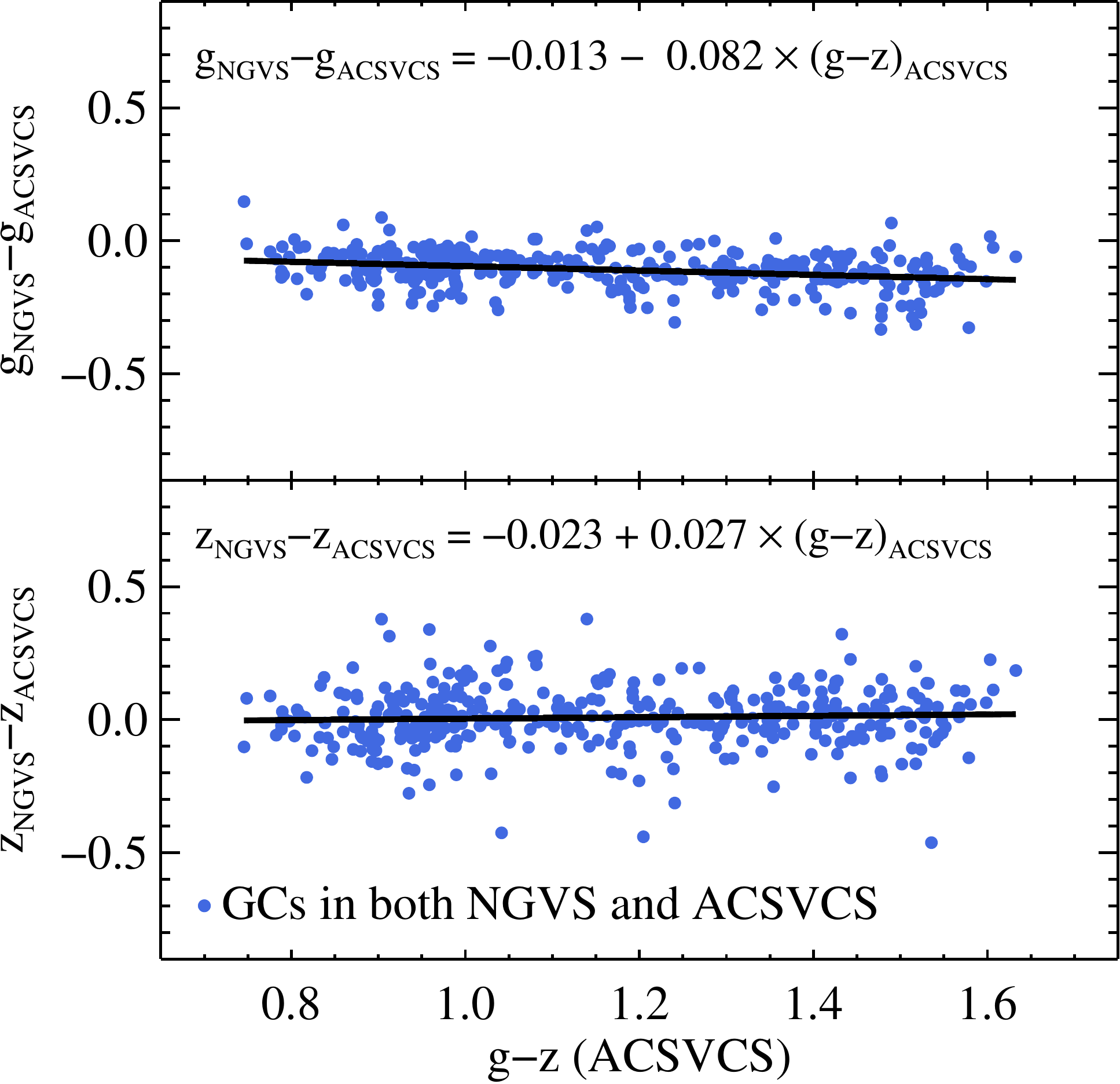}
\caption{The transformation between ACSVCS and NGVS for the $g$ band (upper panel) and $z$ band (lower panel). The blue dots are GC candidates that were detected in both ACSVCS and NGVS around M87 and M49. The black line in each panel shows the best linear fit for the blue dots. The transformation equations adopted in this study are shown in the top of the panels. }
\label{fig:calibration}
\end{figure}

To study the GC systems over a large range in radii, we combine the data from ACSVCS (central region) and NGVS (surrounding area). The details of GC selection in ACSVCS have been well described in \citet{2006ApJ_639_95Peng} and \citet{2009ApJS_180_54Jordan}. For NGVS data, \citet{2014ApJS_210_4Munoz} have described GC selection using the $uiK_s$ color-color diagram. By adding minimal information on the morphology of GCs to the colors, one significantly reduces contamination by foreground stars and background galaxies \citep[][]{2015ApJ_812_34Liu, 2016ApJS_227_12Powalka, 2017ApJ_835_184GonzalezLopezlira, 2020ApJS_250_17Liu, 2020ApJ_899_140Voggel}. In the present study, all the results are based on samples obtained with methods that exploit colors and at least one morphological criterion jointly. For the M87 region, Peng et al. (in preparation) extended the sample of \citet{2016ApJS_227_12Powalka} to fainter GC candidates; for the M49 region, Jain et al. (in preparation) exploited new near-IR data to produce a GC candidates sample with similar criteria.

As discussed in \citet{2014ApJS_210_4Munoz}, the selected GC sample is very clean when the uncertainty of magnitude is smaller than 0.05 mag, which corresponds to $g \sim 24$ mag for NGVS data. In this study, to make the combined GC sample as homogeneous as possible, we adopt an additional magnitude cut of $g<23.9$~mag for both ACSVCS and NGVS GCs, which is the turnover magnitude of GC luminosity function in ETGs \citep{2007ApJS_171_101Jordan, 2010ApJ_717_603Villegas}. In fact, the depth of $K_s$ band data affects our GC selection. At constant $g$ band magnitude, bluer GCs are always fainter in $K_s$ band. So there will be a bias toward selecting slightly redder GCs. We measure the color gradients using optical-selected GC sample (no bias) and optical-$K_s$-selected GC sample. The difference between the two measurements is quite small. That means the selection bias happens at all radii, so may not affect the gradient measurements.

According to previous studies \citep{2014ApJ_794_103Durrell, 2016MNRAS_455_820Oldham, 2017ApJ_835_212Ko}, the GC number density profile changes slope at $\sim 50'-60'$ for M87, and $\sim 45'-60'$ for M49, which corresponds to $\sim 15 R_e$ for these two galaxies. So, we measure the color gradients of GC systems in the range from 0.1 to 15 $R_e$ to remove the effect at the galaxy center and of intra-cluster GCs. The GCs around satellite galaxies and bright stars are also eliminated from the final sample.

The ACSVCS imaging was taken in two filters: $F475W$ ($\approx$ SDSS $g$) and $F850LP$ ($\approx$ SDSS $z$), which are slightly different from the CFHT/Megacam $g$ and $z$ bands used for the NGVS. We use M87 and M49 GC candidates that appear in both ACSVCS and NGVS samples to calculate the transformation between ACSVCS ($g_{\rm ACSVCS}$, $z_{\rm ACSVCS}$) and NGVS ($g_{\rm NGVS}$,  $z_{\rm NGVS}$) magnitudes, as shown in figure~\ref{fig:calibration}. The following equations are adopted to transform the magnitudes between ACSVCS and NGVS:
\begin{equation}
    \begin{aligned}
      & g_{\rm NGVS} = g_{\rm ACSVCS} - 0.013 - 0.082 \times (g-z)_{\rm ACSVCS} \\
      & z_{\rm NGVS} = z_{\rm ACSVCS} - 0.023 + 0.027 \times (g-z)_{\rm ACSVCS}.
    \end{aligned}
    \label{eq:transformation}
\end{equation}
Following \citet{2011ApJ_728_116Liu}, the radial color gradient is defined as:
\begin{equation}
 G_{g-z}=\frac{\Delta (g-z)_0}{\Delta \log R},
\end{equation}
where $(g-z)_0$ is Galactic extinction corrected color index and $R$ is the projected distance from galaxy center. 

\section{RESULTS AND DISCUSSION}
\label{sec:results}

\subsection{The gradients of GC systems}

\begin{figure}
\epsscale{1.15}
\plotone{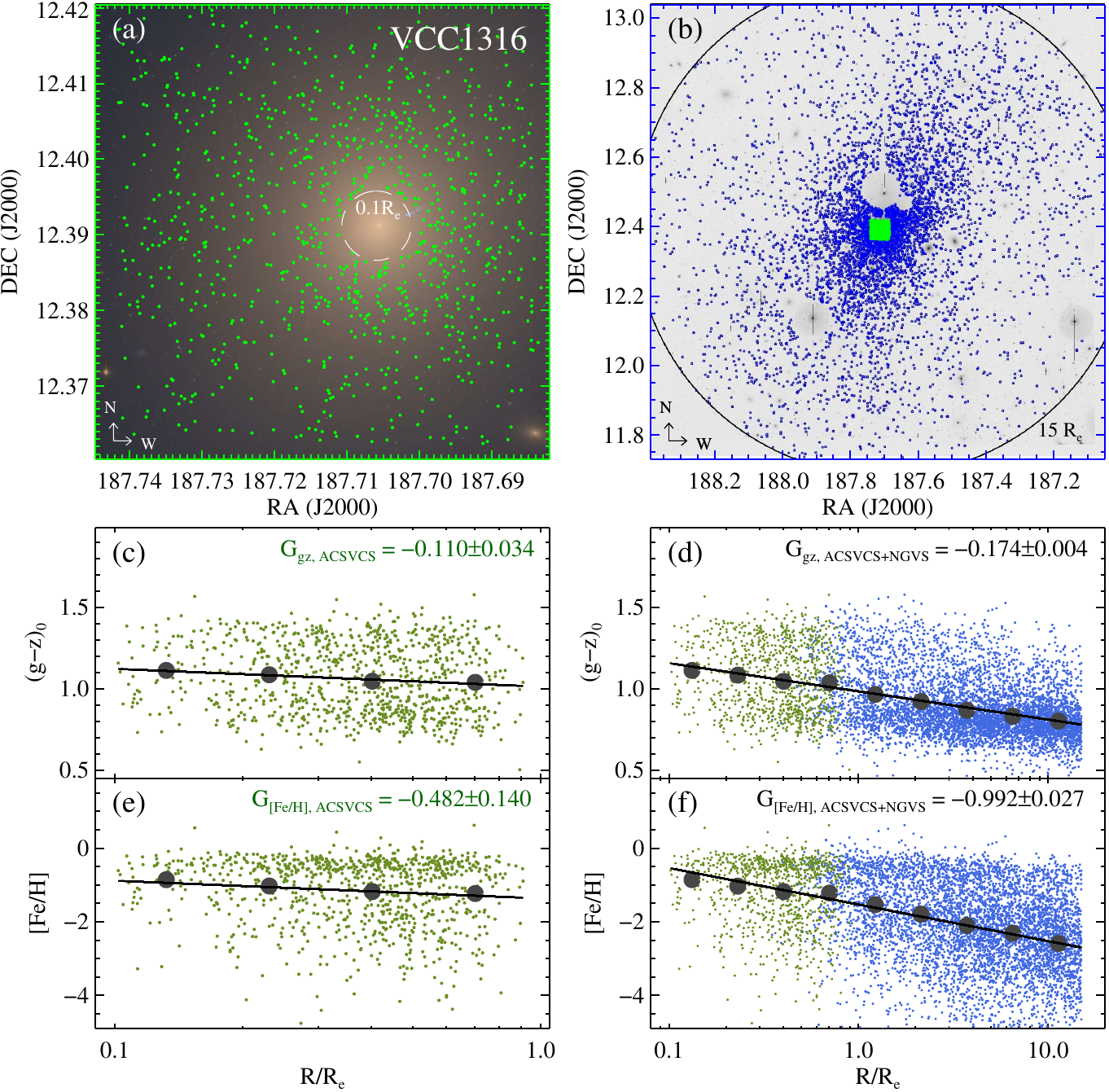}
\caption{\textit{Panel a}: The central region of M87 (VCC1316) and the spatial distribution of ACSVCS GCs (green dots). The dashed green circle is $0.1 R_{\rm e}$. \textit{Panel b}: The spatial distribution of ACSVCS GCs (green dots) and NGVS GCs (blue dots). The black solid circle is $15 R_{\rm e}$. \textit{Panel c}: Color profile of ACSVCS GCs. \textit{Panel d}: Color profile of ACSVCS GCs (green dots) and NGVS GCs (blue dots). \textit{Panel e}: Metallicity profile of ACSVCS GCs. \textit{Panel f}: Metallicity profile of ACSVCS GCs (green dots) and NGVS GCs (blue dots). The black filled circles in the lower four panels are the mean values of $(g-z)_0$ or [Fe/H] in given bins and the black lines are the best-fit values for all the data points.}
\label{fig:gc_all_m87}
\end{figure}

Panels $a$, $c$ and $e$ of Figure~\ref{fig:gc_all_m87} show the spatial distribution, $(g-z)_0$ and [Fe/H] profiles for ACSVCS GCs (green dots) that are in the central region ($\lesssim 1 R_e$) of M87, respectively. Meanwhile, panels $b$, $d$ and $f$ show the plots for ACSVCS (green dots) and NGVS GCs (blue dots) that cover a much larger region ($< 15 R_e$). We measure the mean values of $(g-z)_0$ and [Fe/H] in different radius bins and show these as the black filled circles in the lower four panels. The color gradient of GC system in the central region (ACSVCS GCs only, panel $c$) is $-0.110 \pm 0.034$, which means that the GCs are bluer as one goes outwards. However, the color gradient for the whole GC system (ACSVCS+NGVS GCs, panel $d$) is much steeper ($-0.174 \pm 0.004$). As can be seen in Figure~\ref{fig:gc_all_m87}(d), the number of red and blue GCs are comparable in the central region \citep[see also][]{2006ApJ_639_95Peng} but GCs are dominated by the blue sub-population in the larger radii. Such a distribution of red and blue GCs will cause a steeper color gradient for the whole GC system \citep[e.g.,][]{2006ApJ_636_90Harris, 2008ApJ_681_1233Wehner, 2017MNRAS_470_3227Caso, 2018MNRAS_479_4760Forbes, 2019ApJ_872_202Ko, 2020MNRAS_492_4313DeBortoli, 2021ApJ_915_83Taylor}.

\begin{figure}
\epsscale{1.15}
\plotone{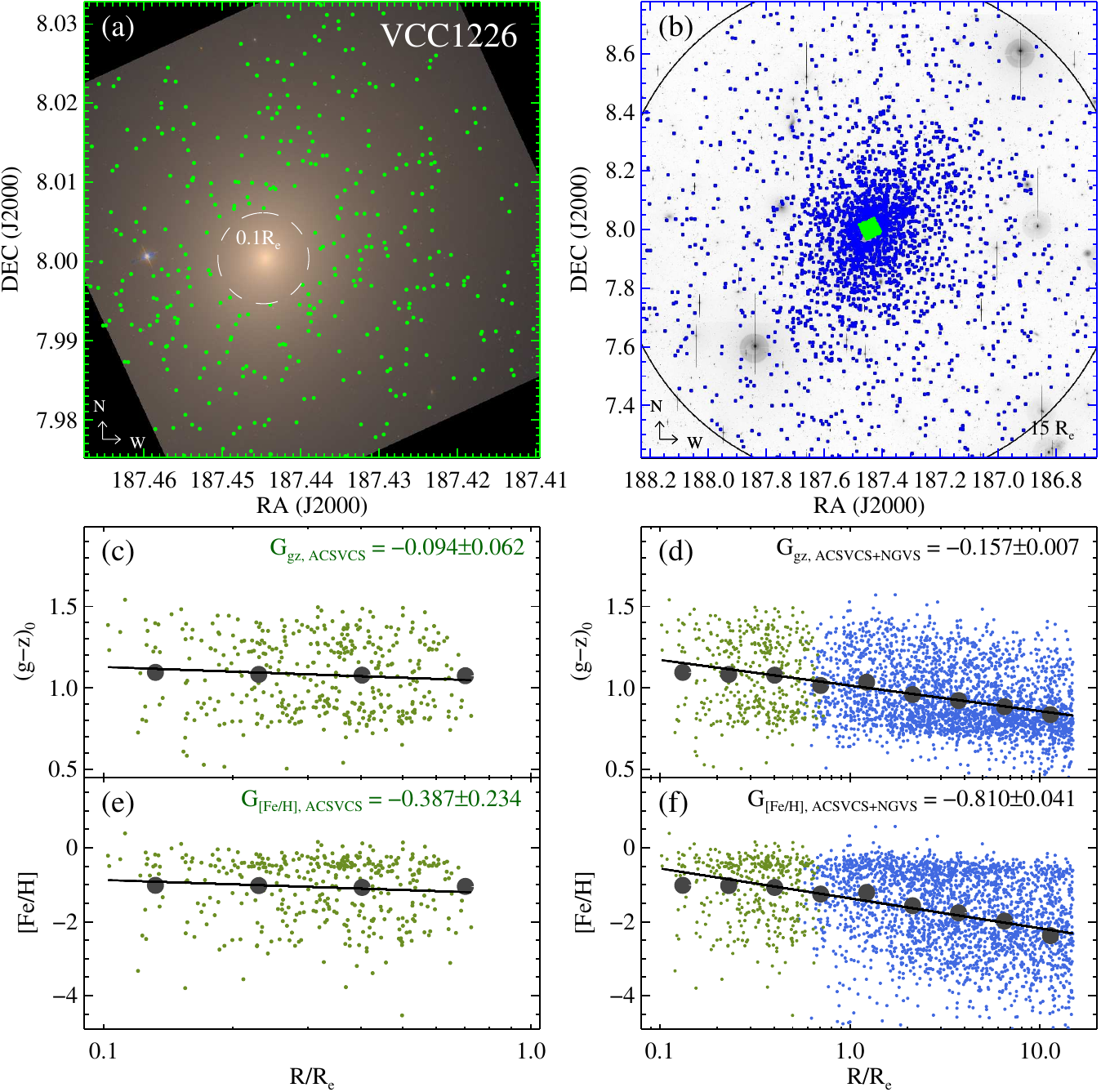}
    \caption{The same as Figure~\ref{fig:gc_all_m87}, but for M49 (VC1226).}
    \label{fig:gc_all_m49}
\end{figure}

Besides the color gradients, we convert the color index to metallicity [Fe/H] according to the nonlinear color-metallicity relation in \citet{2010ApJ_710_51Blakeslee} and calculate the metallicity gradient ($G_{\rm [Fe/H]} \equiv \Delta {\rm [Fe/H]/\Delta \log} R$) of the GC system. We show the [Fe/H] profiles for GC systems of M87 in Figure~\ref{fig:gc_all_m87}(e) and Figure~\ref{fig:gc_all_m87}(f) . The [Fe/H] gradient in the central region is $-0.482 \pm 0.140$ and the gradient for the whole GC system is $-0.992 \pm 0.027$. It is noteworthy that both the color and metallicity gradients are flatter in the central region of M87.

We also investigate the GC system of another massive galaxy, M49, which is the most massive galaxy in Virgo cluster and the center galaxy of subcluster B \citep{2014A+A_570_69Boselli}. The results are shown in Figure~\ref{fig:gc_all_m49}. The color and [Fe/H] gradients are $-0.094 \pm 0.062$ and $-0.387 \pm 0.234$ in the central region while the gradients are $-0.157 \pm 0.007$ and $-0.810 \pm 0.041$ for the whole GC systems. Again, the gradients in the central region are quite flat but the gradients for the whole GC systems are much more significant and steeper.

\subsection{The color gradients of GC sub-populations}

\begin{figure*}
\epsscale{0.9}
\plotone{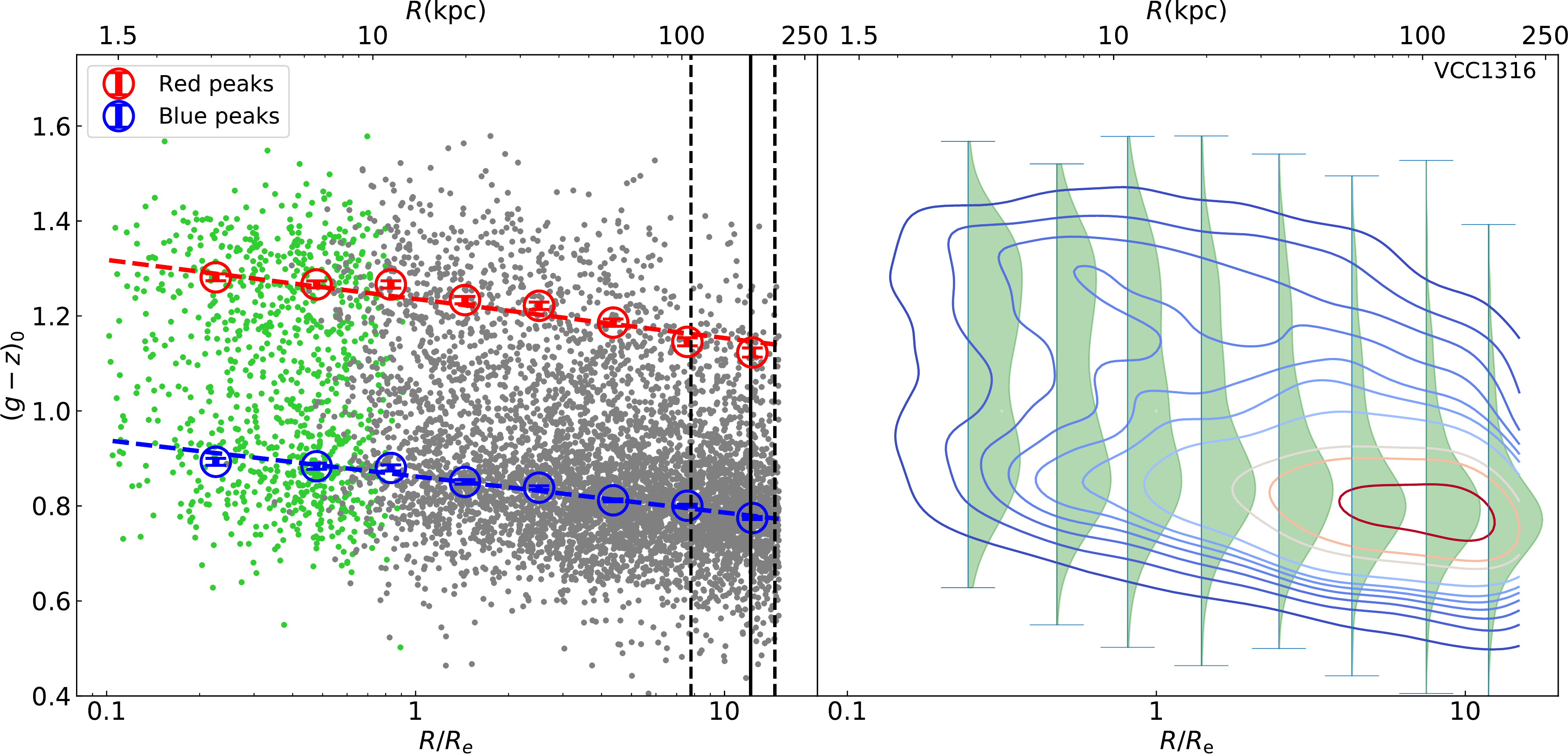}
\caption{The $(g-z)_0$ color profile for M87 GC system. \textit{Left panel:} The data points, including ACSVCS GCs (green dots) and NGVS GCs (gray dots). The black vertical lines mark the transition radius (solid line) and the errors (dashed lines) found in Ko et al. (in preparation). \textit{Right panel:} GC number density map overlaid by color distributions in different bins. The red and blue open circles in the left panel represent the peaks of color distributions of sub-populations. The red and blue dashed lines are the corresponding linear fits.}
\label{fig:gc_subpop_m87}
\end{figure*}

\begin{figure*}
\epsscale{0.9}
\plotone{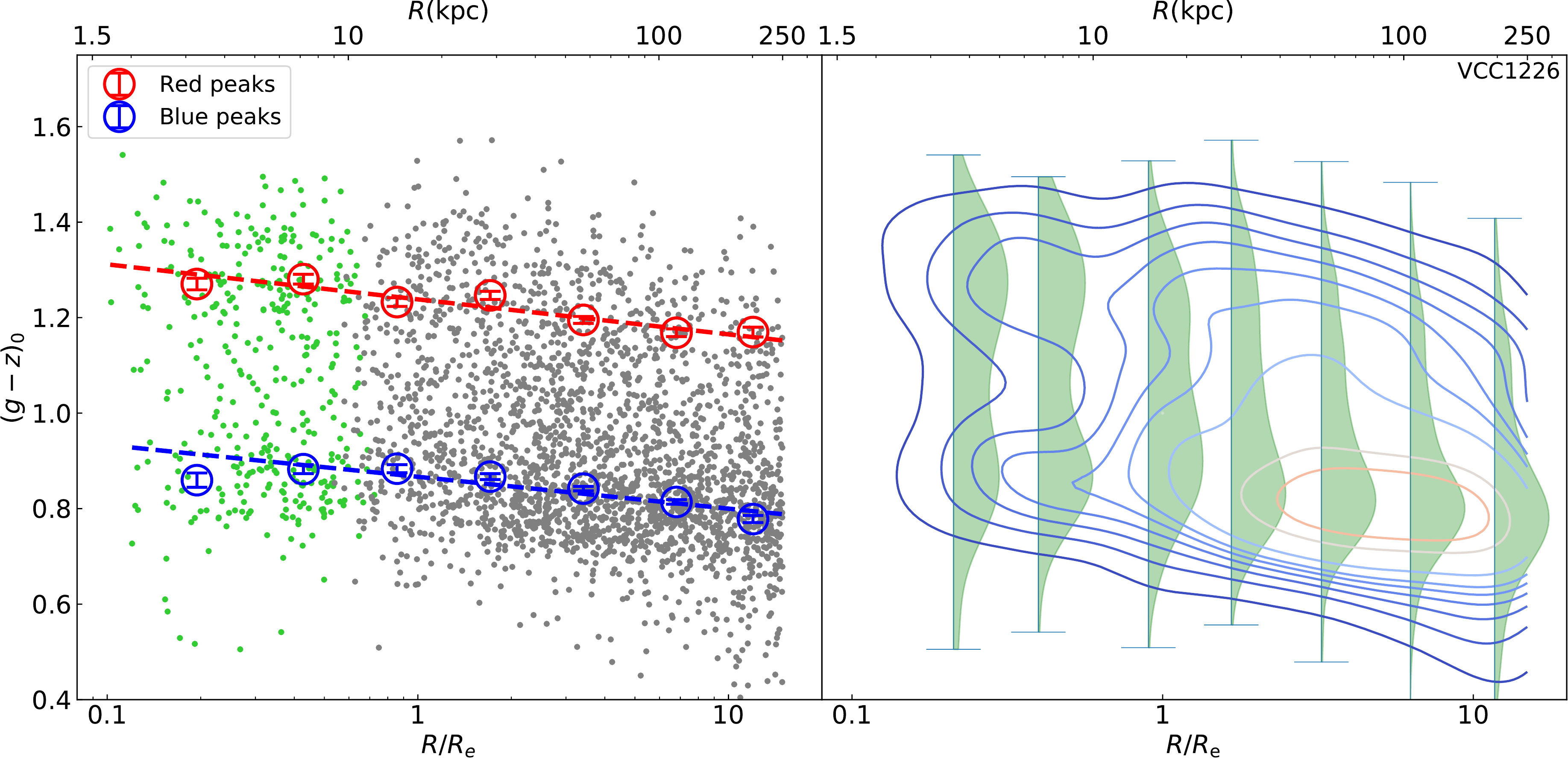}
\caption{The same as Figure~\ref{fig:gc_subpop_m87}, but for M49 GC system.}
\label{fig:gc_subpop_m49}
\end{figure*}

One of the most remarkable properties of GC systems in massive galaxies is their bimodal color distribution, with red and blue sub-populations \citep[e.g.,][]{2006ARA+A_44_193Brodie, 2006ApJ_639_95Peng, 2012A+A_539_54Chies-Santos, 2017ApJ_835_101Harris, 2020MNRAS_492_4313DeBortoli}. It has been widely shown that both red and blue GC sub-populations have negative color/metallicity gradients \citep[e.g.,][]{2009ApJ_703_939Harris, 2011MNRAS_416_155Faifer, 2011ApJ_728_116Liu, 2012MNRAS_421_635Forte, 2014MNRAS_437_273Kartha, 2015ApJ_812_34Liu, 2015MNRAS_451_2625Pastorello, 2018MNRAS_479_4760Forbes}. If so, it is expected that the `dip' of color distribution between red and blue peak will change with galactocentric distance. 

\begin{figure}
\epsscale{1.15}
\plotone{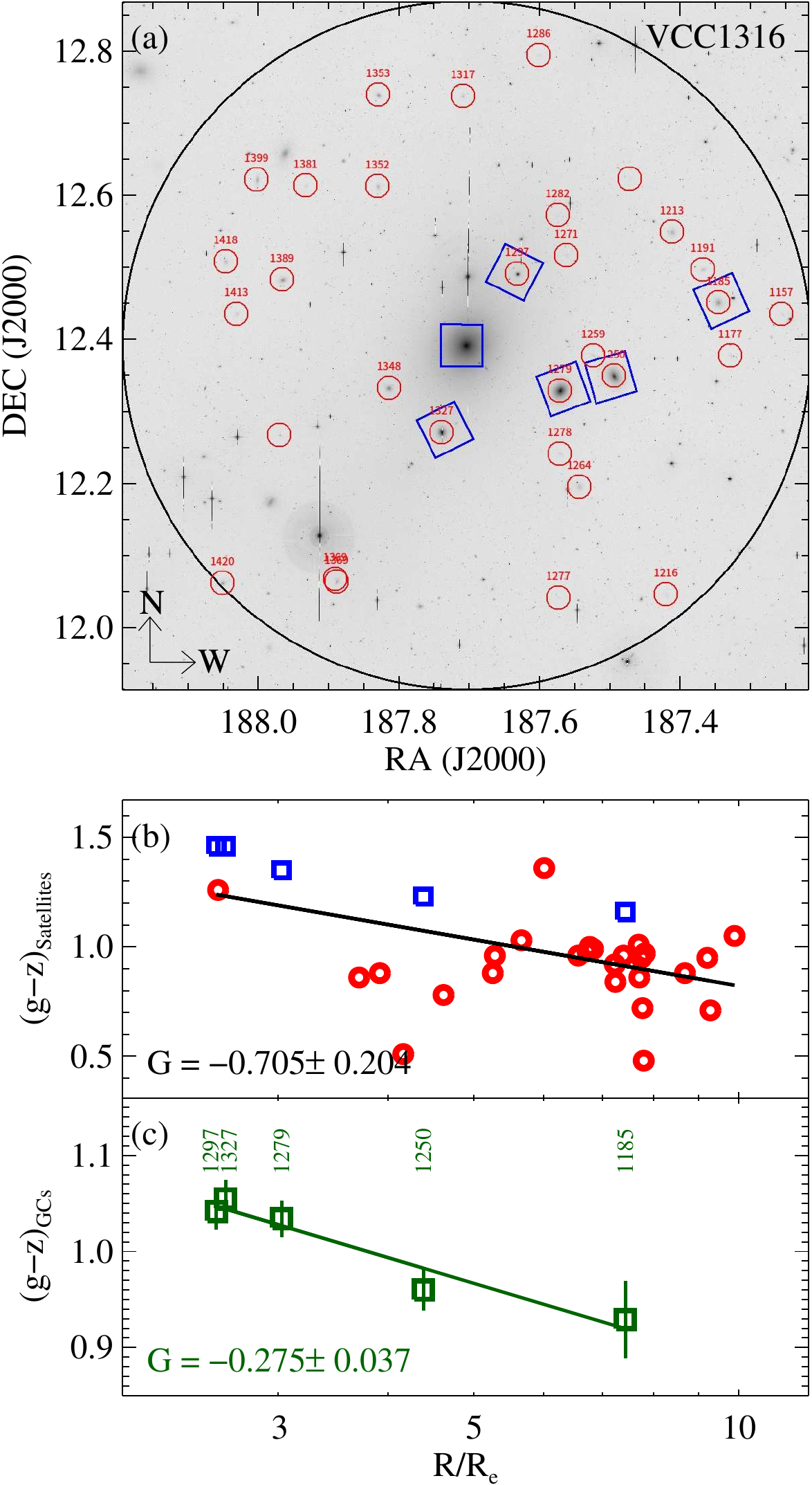}
\caption{\textit{Panel a:} The distribution of satellite ETGs around M87. The ACSVCS galaxies are given by blue squares. The large black circle denotes $15 R_e$ from M87. \textit{Panel b:} The $(g-z)$ color of satellite ETGs as a function of galactocentric radii. \textit{Panel c:} The mean colors of GC systems in satellite galaxies as a function of galactocentric radii.}
\label{fig:dwarf_m87}
\end{figure}

\begin{figure}
\epsscale{1.13}
\plotone{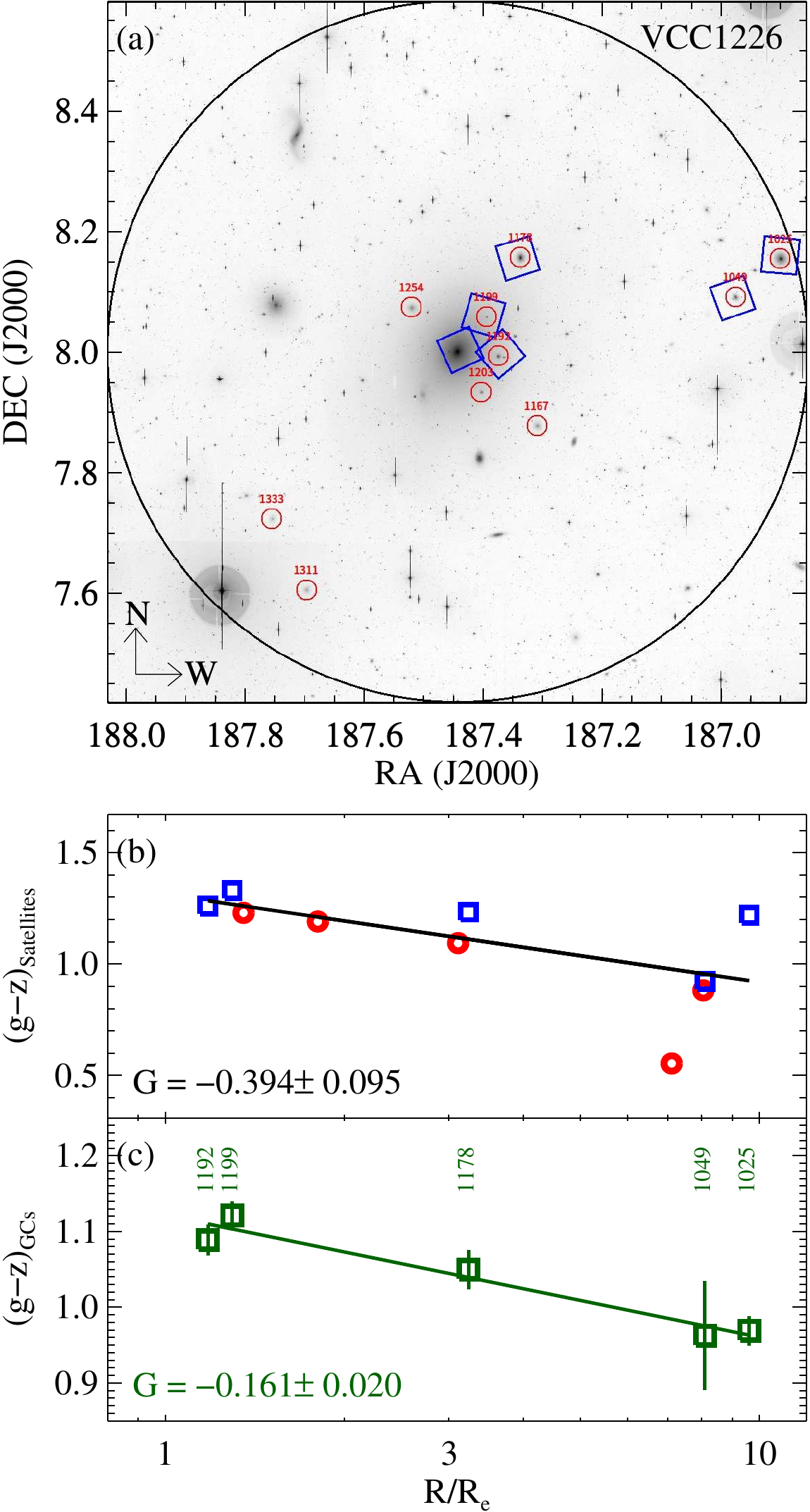}
\caption{The same as Figure~\ref{fig:dwarf_m87}, but for satellite galaxies around M49.}
\label{fig:dwarf_m49}
\end{figure}

Since there are plenty of GCs around M87 and M49, we divide the GCs into several bins according to their galactocentric radii at first, and then we use Kaye's Mixture Model \citep[KMM,][]{1988mmia.book_McLachlan, 1994AJ_108_2348Ashman} to divide the GCs in each radii bins into two sub-populations in the homoscedastic mode, as shown in Figure~\ref{fig:gc_subpop_m87} (GC system of M87) and Figure~\ref{fig:gc_subpop_m49} (GC system of M49). The red and blue open circles in the figures represent the peaks of the color distributions for red and blue sub-populations respectively. The color of both red and blue sub-populations decline with galactocentric radius continually. The typical colors in different radius bins are almost in a straight line for either red or blue sub-population. So we adopt linear fits to the data points and calculate the color gradients for the two sub-populations. The color gradients are $G_{\rm red} = -0.082 \pm 0.005$ and $G_{\rm blue} = -0.076 \pm 0.003$ for M87 GC system, and $G_{\rm red} = -0.073 \pm 0.007$ and $G_{\rm blue} = -0.067 \pm 0.005$ for M49 GC system. All the gradients are negative and significant. The gradients of blue and red GCs in these two massive ETGs are comparable, which is consistent with previous studies \citep{2009ApJ_699_254Harris, 2011ApJ_728_116Liu, 2018MNRAS_479_4760Forbes}.

Many studies found a transition in the color/metallicity profile at several effective radii, beyond which the profiles start to flatten, i.e., there are no longer any color gradients \citep[e.g.,][Ko et al. in prep.]{2009ApJ_703_939Harris, 2011MNRAS_416_155Faifer, 2011MNRAS_413_2943Forbes, 2011ApJS_197_33Strader, 2018MNRAS_479_4760Forbes, 2021ApJ_915_83Taylor}. However, we do not find such a transition radius in M87 and M49. The color peaks of red and blue sub-populations continue to decline up to $\sim 15 R_e$ ($\sim 200 $ kpc for M87 and $\sim 250 $ kpc for M49). There are several possible reasons for this discrepancy. One reason is that each galaxy has its own assembly history, which is related to its mass and the environment. The transition radius is not a universal property for all the galaxies \citep{2101.12216}. Another possibility is the purity of the GC sample. As described in previous studies \citep[e.g.][]{2014ApJS_210_4Munoz, 2018ApJ_864_36Longobardi, 2015ApJ_812_34Liu, 2020ApJS_250_17Liu}, the contamination of the GC sample in Virgo is mainly due to the passive background galaxies. With the increase of the contamination at larger radii, the color profile of GC system could be getting flat. There is another reason, the definitions of color gradient are different in different studies. There are mainly two different definitions of color gradient in previous studies: $\Delta {\rm CI} / \Delta \log R$ and $\Delta  {\rm CI} / \Delta R$, where CI is the color index and $R$ is the galactocentric radius. As shown in Figure~\ref{fig:gc_subpop_m87} and \ref{fig:gc_subpop_m49}, the color profile is almost a straight line in CI-$\log R$ diagram. This is the main reason why we define the color gradient as $G = \Delta {\rm CI} / \Delta \log R$ in this study. If we show the color profile in the  CI-$R$ diagram, the straight line will become a curve, with a steeper slope in the inner region and a shallower slope in the outer region, which seems like a transition.

Ko et al (in preparation) investigate the metallicity gradients for spectroscopically confirmed GCs ($g \lesssim 21.5$ mag) around M87 in a very large radii range ($8 \sim 837$ kpc) and find a transition radius at $169^{+33}_{-60}$ kpc, beyond which the metallicity gradient of blue GCs turns flat. We show their transition radius (vertical solid line) and the errors (two vertical dashed lines) in the left panel of Figure~\ref{fig:gc_subpop_m87}. We can see that the transition radius is close to the boundary of our GC sample. Due to the limitation of our photometric GC sample in this study, we do not rule out the existence of a transition radius.

\subsection{The color gradients of satellite galaxies}

According to the ``two phase" galaxy formation scenario, massive ETGs accrete stars from smaller satellite galaxies in the second phase. In this process, GCs (usually blue) in satellite galaxies are also accreted by the host galaxy. This is one of the formation scenarios of blue GC sub-population in massive ETGs \citep[e.g.,][]{1998ApJ_501_554Cote, 2006ARA+A_44_193Brodie, 2017ApJ_835_101Harris, 2021ApJ_914_20Kang}. Such a formation scenario raises a question. Why do these accreted GCs have a negative color gradient? 

\citet{2016ApJ_818_179Liu} investigated the [$\alpha$/Fe] abundance ratio of 11 low mass ETGs with a large range of cluster-centric distance in the Virgo cluster. They found that the galaxies  closer to M87 (cluster center) tend to have higher [$\alpha$/Fe]. Their results imply that the environment is the key factor for the star and GC formation of low mass ETGs in the galaxy cluster. We select 31 satellite ETGs around M87 ($< 15 Re$) and 10 satellite ETGs around M49 ($< 15 Re$) based on the latest NGVS galaxy catalog \citep{2020ApJ_890_128Ferrarese}. We do not adopt a stellar mass criterion, but all the selected galaxies are certain members of the Virgo cluster, i.e., \texttt{class=1} in table 4 in \citet{2020ApJ_890_128Ferrarese}. Figure~\ref{fig:dwarf_m87} shows the spatial distribution (panel \textit{a}) and color profile (panel \textit{b}) of 31 satellite ETGs around M87. Similar to the result of \citet{2016ApJ_818_179Liu}, the galaxies closer to M87 tend to be redder.  \citet{2006ApJ_639_95Peng} found a positive correlation between the median colors of GC system and the colors of their host galaxies (see Figure 8 in their paper), i.e., the GCs are usually redder in redder host galaxies. We calculate the mean colors of GC systems in 5 ACSVCS galaxies and show them as a function of galactocentric radii in Figure~\ref{fig:dwarf_m87}(c). As expected, the mean colors of GC systems in satellite galaxies decline with galactocentric distance. Figure~\ref{fig:dwarf_m49} shows 10 satellite ETGs around M49, including 5 ACSVCS galaxies. The results of M49 satellite ETGs are the same as those of M87 satellite ETGs. 

For massive ETGs like M87 and M49, the central regions are dominated by the \textit{in situ} stars/GCs from compact progenitor, which are formed at higher redshifs and thus with redder colors, and the outer regions are dominated by the \textit{ex situ} stars/GCs accreted by the satellite galaxies \citep[e.g.,][]{2019MNRAS_487_318Karademir, 2021A+A_647_95Pulsoni}. The existence of the color gradient of the satellite galaxy system can explain the color gradients of the GC system around massive ETGs under the framework of ``two phase" galaxy formation scenario. Early stage major mergers mainly influence the distribution of GCs in the central region and flatten pre-existing gradients \citep{2009A+A_499_427DiMatteo, 2012MNRAS_421_2478Tortora, 2019ApJ_880_111Oyarzun}, while the following successive minor mergers in the second phase provide GCs to the outer regions. As implied in the Figures~\ref{fig:dwarf_m87}(c) and \ref{fig:dwarf_m49}(c), the GCs accreted from these satellite galaxies naturally have a negative color gradient. 

\section{SUMMARY}	
\label{sec:summary}

With the combination of GC samples from ACSVCS and NGVS, we mainly investigate the color gradients of GC systems in M87 and M49 over a very large range of radii, from the central region to $15\,R_e$ ($\sim 200 $ kpc for M87 and $\sim 250 $ kpc for M49). We find that:

\begin{enumerate}

\item The color and metallicity of the whole GC system continues to decline with the galactocentric radii, from 0.1 to 15 $R_e$. The gradient is shallower in the central region ($\lesssim R_e$).

\item Both the red and blue GC sub-populations show significant negative color gradients over a large range of galactocentric radii ($0.1 R_e$ to $15 R_e$). We do not find any sign of transition in the color profiles.

\item The satellite ETGs closer to the center of the host galaxy tend to be redder, as are their GC systems. In an accretion scenario, this may naturally explain the color gradients seen in the GC systems of massive galaxies.

\end{enumerate}

\acknowledgments

The authors acknowledge support from the National Natural Science Foundation of China (NSFC, Grant No. 12173025, 11673017, 11833005, 11933003, 11890692, 11621303). This work is also supported by the Shanghai Natural Science Foundation (Grant No. 15ZR1446700), 111 project (No. B20019), Shanghai Key Laboratory for Particle Physics and Cosmology (SKLPPC), and Key Laboratory for Particle Physics, Astrophysics and Cosmology, Ministry of Education. We also acknowledge the science research grants from the China Manned Space Project with NO. CMS-CSST-2021-A02.

Facility: HST, CFHT


\bibliographystyle{aasjournal}

\end{CJK*}
\end{document}